\documentclass[conference]{IEEEtran}
\IEEEoverridecommandlockouts
\usepackage{cite}
\usepackage{amsmath,amssymb,amsfonts}
\usepackage{graphicx}
\usepackage{textcomp}

\usepackage{comment}
\usepackage{mathtools} %
\usepackage{todonotes} %
\usepackage[caption=false]{subfig} %
\usepackage[mathscr]{euscript} 
\usepackage{url} %
\usepackage{hyperref}
\usepackage{bm}
\usepackage{algpseudocode} 
\usepackage{algorithm}
\usepackage{xifthen}
\usepackage[export]{adjustbox}
\usepackage{amsmath}
\usepackage{listings}
\usepackage{caption}
\usepackage{xcolor}

\usepackage{amsmath}
\usepackage{graphicx}
\usepackage{subfig}
\usepackage{wrapfig}
\usepackage{array}

\usepackage{colortbl}  

\usepackage{algorithm}
\usepackage{algpseudocode}
\usepackage{tikz}
\usetikzlibrary{shapes, positioning, arrows.meta}
\usepackage{varwidth}
\usepackage{booktabs} 
\definecolor{mygreen}{rgb}{0,0.6,0}
\definecolor{mygray}{rgb}{0.5,0.5,0.5}
\definecolor{mymauve}{rgb}{0.58,0,0.82}

\lstdefinelanguage{cypher}{
    sensitive=true,
    morekeywords=[1]{MATCH, RETURN, WHERE},
    morekeywords=[2]{PERSON, FRIEND},
    morestring=[b]",
    morecomment=[l]{//},
    morecomment=[s]{/*}{*/},
    morecomment=[s]{--}{\ },
}

\lstdefinestyle{cypherstyle}{
    language=cypher,
    basicstyle=\small\ttfamily,
    keywordstyle=[1]\color{blue},
    keywordstyle=[2]\color{red},
    commentstyle=\color{mygreen},
    stringstyle=\color{mymauve},
    numberstyle=\tiny\color{mygray},
    breaklines=true,
    showstringspaces=false,
    captionpos=b
}

\usepackage{float}
\usepackage{subcaption} 
\usepackage{svg}
\newcommand{\mycomment}[1]{}
\NewDocumentCommand{\vect}{ O{} O{} m }{\mathbf{#3}\ifthenelse{\isempty{#1}}{}{^{(#1)}}\ifthenelse{\isempty{#2}}{}{_{#2}}}

\NewDocumentCommand{\mat}{ O{} O{} m }{\mathbf{#3}\ifthenelse{\isempty{#1}}{}{^{(#1)}}\ifthenelse{\isempty{#2}}{}{_{#2}}}

\NewDocumentCommand{\ten}{ O{} O{} m }{\pmb{\mathscr{#3}}\ifthenelse{\isempty{#1}}{}{^{(#1)}}\ifthenelse{\isempty{#2}}{}{_{#2}}}

\def\BibTeX{{\rm B\kern-.05em{\sc i\kern-.025em b}\kern-.08em
    T\kern-.1667em\lower.7ex\hbox{E}\kern-.125emX}}

\makeatletter
\renewcommand{\ALG@beginalgorithmic}{\scriptsize}
\makeatother
\begin{document}
\title{Binary Bleed: Fast Distributed and Parallel Method for Automatic Model Selection}

\author{\IEEEauthorblockN{
Ryan Barron\IEEEauthorrefmark{2}\IEEEauthorrefmark{3},
Maksim E. Eren\IEEEauthorrefmark{1}\IEEEauthorrefmark{3}, 
Manish Bhattarai\IEEEauthorrefmark{2}, \\
Ismael Boureima\IEEEauthorrefmark{2}, 
Cynthia Matuszek\IEEEauthorrefmark{1}\IEEEauthorrefmark{3},
and Boian S. Alexandrov\IEEEauthorrefmark{2}
}
\IEEEauthorblockA{
\IEEEauthorrefmark{2}Theoretical Division, Los Alamos National Laboratory. Los Alamos, USA. \\
\IEEEauthorrefmark{1}Advanced Research in Cyber Systems, Los Alamos National Laboratory. Los Alamos, USA. \\
\IEEEauthorrefmark{3}Department of Computer Science and Electrical Engineering, University of Maryland, Baltimore County. Maryland, USA.}
\thanks{U.S. Government work not protected by U.S. copyright.}
}

\maketitle
\vspace{-40em}

\begin{abstract}
In several Machine Learning (ML) clustering and dimensionality reduction approaches, such as non-negative matrix factorization (NMF), RESCAL, and K-Means clustering, users must select a hyper-parameter $k$ to define the number of clusters or components that yield an ideal separation of samples or clean clusters. This selection, while difficult, is crucial to avoid overfitting or underfitting the data. 
Several ML applications use scoring methods (e.g., Silhouette and Davies Boulding scores) to evaluate the cluster pattern stability for a specific $k$. The score is calculated for different trials over a range of $k$, and the ideal $k$ is heuristically selected as the value before the model starts overfitting, indicated by a drop or increase in the score resembling an elbow curve plot.
While the grid-search method can be used to accurately find a good $k$ value, visiting a range of $k$ can become time-consuming and computationally resource-intensive.
In this paper, we introduce the \textit{Binary Bleed} method based on binary search, which significantly reduces the $k$ search space for these grid-search ML algorithms by truncating the target $k$ values from the search space using a heuristic with thresholding over the scores. Binary Bleed is designed to work with single-node serial, single-node multi-processing, and distributed computing resources. In our experiments, we demonstrate the reduced search space gain over a naive sequential search of the ideal $k$ and the accuracy of the Binary Bleed in identifying the correct $k$ for NMFk, K-Means pyDNMFk, and pyDRESCALk with Silhouette and Davies Boulding scores. We make our implementation of Binary Bleed for the NMF algorithm available on GitHub.
\end{abstract}

\begin{IEEEkeywords}
K search, Optimization, Binary Search, Log Search, Binary Chop, Parameter Search
\end{IEEEkeywords}
\maketitle

\section{Introduction}
\label{sec:introduction}
In the machine learning (ML) landscape, clustering, and dimensionality reduction are critical for revealing underlying data patterns, simplifying complex data, and improving predictive models' performance. Methods such as non-negative matrix factorization (NMF), K-means, pyDNMFk, and pyDRESCALk clustering are popular because they effectively handle large data. However, these techniques depend heavily on an appropriate number of clusters or components, denoted by the hyper-parameter $k$. This parameter setting is essential as it influences the ability of the model to capture the intrinsic data structure without overfitting or underfitting accurately.

One approach to determine the NMF's optimal $k$ is automatic model determination (NMFk), a heuristic algorithm based on silhouette scores to identify a $k$ to yield stable patterns in $\mat{W}$ and clusters in $\mat{H}$ \cite{alexandrov2020patent, SmartTensors, TELF}. Here, $k$ is the highest $\mat{W}$ and $\mat{H}$ silhouette scores above the given thresholds $t_W$ and $t_H$. In NMFk, this selection of $k$ is done by a grid search over a $k$ search space $k \in \mathcal{K}=\{1,2,...,K\}$ defined by the user, where a sudden drop in the silhouette scores after the ideal $k$ value indicates over-fitting. Such selection of the $k$ hyper-parameter is common in other ML approaches, such as K-means Clustering or RESCAL clustering~\cite{bhattarai2023distributed}. Various distributed versions of unsupervised techniques have been specifically developed to manage large-scale datasets effectively, as detailed in the literature \cite{boureima2022distributed,boureima2024distributed,bhattarai2020distributed,bhattarai2023distributed,BHATTARAI2023104709}. While these advancements represent significant progress, the implementations' time complexity remains a critical issue. This complexity is primarily influenced by the parameter $k$ due to the requirement to evaluate cluster stability across a range of $k$ values. Consequently, there is a pressing need to develop an improved selection algorithm that can effectively narrow down the search space for $k$, thereby optimizing the computational efficiency of these techniques.

This work introduces a method based on binary search, \textit{Binary Bleed}, which heuristically prunes the search space for the hyper-parameter $k$. Instead of visiting each $k$ sequentially, Binary Bleed performs a binary search over the given $k$ search space. 
It operates under the assumption that the clustering method is used with a scoring metric that increases with $k$ for all sub-optimal $k$ values while remaining low for $k>k_{optimal}$ due to over/under-fitting. This assumption has been valid for most clustering and dimensionality reduction algorithms, including NMFk and RESCALk when paired with silhouette scoring and K-means when paired with Davies Boulding scoring.  
Practically, in NMFk, this translates into silhouette scores approximating the shape of a square wave curve.
While the working assumption has been tested and validated against various datasets, including text data for topic modeling, cyber-security data for malware analysis and anomaly detection ~\cite{10.1145/3558100.3563844, 10527250, 10.1145/3624567, SmartTensors}, the authors acknowledge the potential it may not apply to other datasets. 
 
Based on the assumptions discussed above, Binary Bleed performs binary search over the $k$ search space $\mathcal{K}$, given user-provided threshold $t$, and prunes the lower $k$ from $\mathcal{K}$ on identifying a score $s>=t$ for maximization tasks and $s<=t$ for minimization tasks. 
In the best-case scenarios, Binary Bleed drastically reduces the $k$ search space, allowing for faster hyper-parameter tuning for $k$ selection. We demonstrate $k$ visit reductions with Binary Bleed when finding the true $k$ compared to a naive linear search of $k$ while preserving the correct identification of $k$. The method is analyzed with NMFk, K-means, pyDNMFk, and pyDRESCALk algorithms using silhouette scoring for the maximization task and Davies Boulding scoring for the minimization task. We further develop Binary Bleed to work with multi-processing and High-Performance Computing (HPC) systems and make the algorithm available on GitHub\footnote{NMFk with Binary Bleed is available in \url{https://github.com/lanl/T-ELF}.}. To the best of our knowledge, integrating binary search with a pruning heuristic for $k$ selection with scoring tending towards square-wave distributions is novel.

\section{Relevant Work}
\label{sec:relevant_work}
This section briefly reviews several machine learning dimensionality reduction and clustering algorithms that require hyper-parameter selection for $k$. Binary Bleed uses binary search to optimize $k$, so we review related binary search studies. Finally, Binary Bleed optimizes the hyper-parameter $k$ over a search range, making an automatic model determination, or $k$ selection, another foundation. Other works optimized their respective algorithms over parallel and distributed contexts for binary search and automatic model determination.

\textbf{Applications of $k$ Search}: Optimizing for $k$ has utility in several domains, particularly clustering, where optimal cluster count search computations can be expensive. Typically, $k$ is a user-provided parameter that must be refined through trial and error. Algorithms that require user-specified or trial-discovered $k$ include K-means Clustering \cite{MacQueen1967SomeMF}, K-medoids Clustering \cite{medioids}, K-medians Clustering \cite{jain1988algorithms}, Fuzzy C-means Clustering \cite{BEZDEK1984191}, Mini-Batch K-means \cite{mini_batch}, Spherical K-means \cite{spherical_k_means}, Elkan K-means \cite{elkan2003using}, NMF Clustering \cite{xu2003document}, Symmetric NMF Clustering \cite{symmetric_nmf}, and RESCAL Clustering \cite{nickel2011three}.

Each algorithm can use the silhouette score to determine the ideal $k$ heuristically. In our experiments, detailed in the results section, we analyze Binary Bleed's performance when operating with the silhouette score and Davies Boulding to test NMFk, RESCAL, and K-means algorithms.

\textbf{Binary Search}:
Several aspects of binary search are relevant to this work, including optimizing search, modifying binary search, and parallelizing binary search. Noisy Binary Search \cite{Karp2007NoisyBS} minimizes the number of comparisons to find an optimal coin in a sequence of coin flips by using biased coin flips to compare elements in a sorted sequence indirectly. It identifies the position where the probability of observing heads changes from below to above a given target threshold. Flip positions are determined from previous outcomes, focusing the flips on areas with the highest uncertainty about crossing the target probability. This approach is similar to Binary Bleed, which narrows searches in the search space using a thresholding mechanism combined with an observed score. Differently, while Noisy Binary Search selects the next target based on the score of the current item, Binary Bleed is designed for methods like NMF, where silhouette scores obtained from different $k$ selections are independent of each other.
Consequently, the silhouette score cannot decide the next $k$ to visit. Instead, for the $k$ search space pruning heuristic, Binary Bleed follows the hypothesis that the silhouette score will remain low after the ideal $k$ due to the overfitting phenomenon. Once a stable $k$ is found, all the lower $k$ are ignored, and higher or lower $k$ values require visitation until overfitting is observed.

Another work, \cite{Tyrrell1988}, sought to optimize the binary search algorithm by modifying it to check the bounds of a stack and adjust these bounds, thereby increasing or narrowing the current search space to make the search process more efficient. This was required to measure the resting position of human eyes, which changes. 
Similar to \cite{Tyrrell1988}, a modified binary search was described by \cite{ChadhaMM14}, which checks at both ends of a sub-array and the middle to reduce the number of iterations through many sub-arrays to find the search value. Both \cite{Tyrrell1988} and \cite{ChadhaMM14} reduce needed checks in binary search, similar to this work. However, this work precomputes the search space and iteratively prunes values. Binary Bleed does not terminate upon finding an optimal candidate. Instead, it adjusts the search space and continues exploring the parameter direction to optimize the results further.

Some works have achieved parallelization of the binary search with specific constraints. One work, \cite{efficient_parallel_binary}, partitions the search array into smaller sub-arrays, searching each in parallel, which is similar to \cite{ChadhaMM14} but operates all arrays concurrently, and so the similarities to this work are the same with the addition of parallelizing the problem. Binary Bleed may operate parallel jobs across single or multiple compute nodes, each handling a portion of the $k$ search space. Another work, Parallel Binary Search in \cite{parallel_binary} defines two arrays, $A$ and $B$, where $|A|<|B|$. Parallelization distributes $A$'s elements across resources to find the smallest element in $B$ greater than or equal to each element in $A$. While this approach to parallelization is similar to parallelizing the $k$ range in our work, the objectives differ. Binary Bleed aims to find the largest or smallest optimal value in the $k$ space rather than minimizing array indices.

Importantly, distributed and parallel binary searches are often used interchangeably but address different challenges. In a parallel search, data is small enough to fit in memory, so evaluations of different $k$ values are concurrent on different processors or nodes. In contrast, distributed search requires multiple processors or nodes to handle single $k$ evaluations for data too large for memory.
In our context, comparing the search term to the current term $n$ in binary search is analogous to model computation at $k=n$. This computation at $k=n$ may exceed memory capacity for large datasets, necessitating a distributed solution where the data for a single calculation is divided among multiple resources. We demonstrate this using Binary Bleed to minimize the $k$ search space with distributed NMF from our prior work \cite{pyDNMFk,bhattarai2023distributed}. This approach allows parallel evaluations to concurrently minimize $k$, while distributed evaluations manage the computation for each $k$ in large datasets, ensuring efficient utilization of resources.

\textbf{K Search}:
Several algorithms have optimized parameters or $k$, including parallelization and distributed contexts. 
For textures,  \cite{distributed_constrained_heuristic_search} uses distributed calculations from constraints and backtracking error correction, allowing bulk calculation with minor corrections.  
While our work can distribute an NMF $k$ operation larger than memory, there is no need to backtrack in the $k$ optimization. Either an optimal $k$ is found or not and can be done in parallel and/or distributed calculations. 


In an alternative method to approximate the number of $k$ in a dataset, \cite{distributedNearestNeighbor} uses hash functions with prototypes. This approach adaptively chooses a set of prototypes and binary codes to represent the data sample distribution, approximating neighbor relationships. The distribution in this algorithm is derived from learning the weights of the hash codes rather than selecting $k$. Therefore, while distribution is utilized in their method, our present work directly employs distribution and parallelization to determine the optimal $k$ value.

Another optimization can be seen in the context of K Nearest Neighbors (KNN) \cite{big_data_knn}, where training examples are sorted into a binary search tree based on the two furthest points in the dataset. The furthest data points are identified at each node to construct the next partition. This method attempts to optimize the $k$ selection for KNN. However, it is highly specific to KNN rather than a general parameter search optimization method, as our work with Binary Bleed proposed.

Similar to \cite{big_data_knn}, the parallel $k$ search in \cite{sismanis2012parallel} utilizes GPU-accelerated parallel processing and specific sorting algorithms, such as bitonic sort, that are efficient under synchronous operation conditions. While this approach reduces the number of operations needed in a distributed environment, it is tailored to algorithms that benefit from GPU operations and truncated sorting methods. In contrast, our work aims to optimize the $k$ parameter more generally and does not rely on a specific model, sorting method, or primitive operation.

\section{Algorithm}
\label{sec:algorithm}

Estimating the optimal number of clusters $k$ is crucial for effective unsupervised machine learning methods for latent feature extraction and clustering. Traditional methods use an exhaustive linear search, resulting in high computational costs, scaling with \( \Theta(n) \), where $n$ is $|\mathcal{K}|$. Binary search offers a more efficient alternative with a worst-case time complexity of \( O(\log(n)) \), but it usually finds only exact matches. The Binary Bleed algorithm adapts binary search to optimize clustering performance metrics, like silhouette scores. The new methods introduced are Binary Bleed Vanilla (Vanilla) and Binary Bleed Early Stop (Early Stop), while models not using these methods are referred to as Standard.

\subsection{ Binary Bleed }

Our method is named Binary Bleed due to its method of pruning less optimal \( k \) values, extends traditional binary search techniques by continuing to search even after a potential optimal \( k \) is found, ensuring the maximization or minimization of the scoring function \( f: \mathcal{K} \times \mathcal{D} \rightarrow \mathbb{R} \), where \( \mathcal{K} \) is the set of candidate parameters and \( \mathcal{D} \) represents the dataset.
Binary Bleed operates recursively to identify the optimal \( k \) that maximizes or minimizes an evaluation score function \( f(k, \mathcal{D}) \). Unlike a conventional binary search that terminates upon finding a target value, Binary Bleed dynamically adjusts the search space based on an evaluation threshold. From an initial range \([k_{\text{min}}, k_{\text{max}}]\), it computes the mid-point \[
k_{\text{mid}} = \left\lfloor \frac{k_{\text{min}} + k_{\text{max}}}{2} \right\rfloor
\] and evaluates the score \[
f_{\text{mid}} = f(k_{\text{mid}}, \mathcal{D}).
\] If \( f_{\text{mid}} \) meets the threshold, the search continues in the direction of optimization. For maximization, 

the lower bound is updated to \( k_{\text{min}} = k_{\text{mid}} + 1 \) when $f(k_{\text{mid}} + 1, \mathcal{D}) > f_{\text{mid}}$, 
and conversely, the upper bound is updated to \( k_{\text{max}} = k_{\text{mid}} - 1 \) when  $f(k_{\text{mid}} - 1, \mathcal{D}) < f_{\text{mid}}$. For minimization, the process is reversed. This recursive update continues until convergence or until a predefined stopping criterion is met, allowing the algorithm to ``bleed" into higher or lower \( k \). This ensures thorough exploration and optimization of the parameter \( k \), adapting the search space dynamically to achieve the best possible evaluation score within the specified constraints.

This search does not cease upon finding a \( k \) above the threshold; instead, it continues exploring to ensure that no better solutions exist, particularly focusing on larger \( k \) values that may yield higher scores. The runtime complexity of Binary Bleed is bounded by \(\Theta(n^{\log_2{(p + 1)}} )\times (T_{\text{model}} + T_{\text{score}})\) in both the best and worst cases, as shown later in \ref{runtime}, where \( p \) is the probability of recursing twice. We apply binary search on the ordered set \(\mathcal{K}\), the search space, to find its optimal value. Mathematically, this optimization can be expressed as:
\[
k_{\text{optimal}} = \max \left\{ k \in \{1, 2, \ldots, K\} : S(f(k)) > T \right\}
\]
where \( T \) is the selection threshold, \( f(k) \) represents the model computation, and \( S(f(k)) \) is the model's scoring function.

\begin{algorithm}[t!]
\caption{Binary Bleed $k$ Search, Single Rank \& Thread }\label{alg:binary_search_optimal_k}
\begin{algorithmic}[1]
\Require \( \mathcal{K} \) (list of \( k \) ), $data$ (dataset), $model$ (model for calculation), $scorer$ (function to quantify output), $T_{\text{select\_k}}$ (minimum score for optimal \( k \))
\Ensure Optimal $k$ value and its prediction score
\State $ranks_{\text{seen}} \gets \emptyset,   k_{\text{max}} \gets \infty ,  k_{\text{min}} \gets -\infty$  

\Function{BinaryBleedKSearch}{ $\mathcal{K} ,i_{left}, i_{right}, scorer, data, model, T_{\text{select\_k}}$}
    \If{$i_{left} \geq i_{right}$} \Return \EndIf
\State $middle \gets i_{left} + \left\lfloor \frac{i_{right} - i_{left}}{2} \right\rfloor$
    \State $k_{\text{middle}} \gets Ks[middle]$
    \If{$k_{\text{middle}} > k_{\text{min}}$ \textbf{and} $ k_{\text{middle}} < k_{\text{max}}$} 
    \State $score \gets scorer(model(data, k_{\text{middle}}))$
    \State Append $(k_{\text{middle}}, score)$ to $ranks_{\text{seen}}$
    \If{$score \geq k_{\text{select\_threshold}}$ } $k_{\text{min}} \gets k_{\text{middle}}$
    \EndIf   
    \If{$score \leq k_{\text{stop\_threshold}}$ } $k_{\text{max}} \gets k_{\text{middle}}$
    \EndIf
    \EndIf
    \If{$middle+1 \leq k_{\text{max}}$ }
        \State \Call{BinaryBleedKSearch}{ $Ks, k_{\text{max}},  k_{\text{min}},middle+1, i_{right}$
        
            $ranks_{\text{seen}}, scorer, data, model, T_{\text{select\_k}}$}
    \EndIf
    \If{$middle-1 \geq k_{\text{min}}$ }
        \State \Call{BinaryBleedKSearch}{ $Ks, k_{\text{max}},  k_{\text{min}},i_{left}, middle-1$
        
            $ranks_{\text{seen}}, scorer, data, model, T_{\text{select\_k}}$}
    \EndIf
\EndFunction
\end{algorithmic}
\end{algorithm}

The algorithm is presented in Algorithm~\ref{alg:binary_search_optimal_k}, showcasing its operational structure. To determine the optimal \( k \), a list of \( k \) (\( \mathcal{K} \)) is provided with the dataset, model, scorer, and score threshold. Initially, the set of visited \( k \), the maximum (\( k_{\text{max}} \)), and minimum (\( k_{\text{min}} \)) bounds are initialized (lines 1-2). The algorithm checks the base case for recursion: if the left index \( i_{left} \) is greater than or equal to the right index \( i_{right} \), the function returns, terminating the recursive search when no more \( k \) values are left to explore (lines 3-4).  The optimal \( k \) is sought by computing the middle index of $\mathcal{K}$ and its \( k \) (lines 5-6). This value is validated to ensure it is within  \( k_{\text{min}} \) and  \( k_{\text{max}} \)  (line 7). The model is evaluated at this middle \( k \)  on the dataset, and the score is added to the set of visited \( k \)  values (lines 8-9). If the score at  \( k_{\text{middle}} \) meets or exceeds the threshold, \( k_{\text{min}} \) updates to \( k_{\text{middle}} \), focusing on larger values (lines 10-12). If the score is below the stop threshold, \( k_{\text{max}} \) updates to  \( k_{\text{middle}} \), focusing on lower values (lines 13-15). The BinaryBleedKSearch function is recursively called on the narrowed \( k \) range (lines 17-19). This dynamically recursive search space adjustment allows thorough exploration of \( k \), ensuring the optimal is found based on the evaluation criteria.

\textbf{Recursive Behavior Analysis}:\label{runtime} Let \( T(n) \) denote the runtime of the algorithm for \( \mathcal{K} \) of size \( n \), and $T_{\text{model}}$ and $T_{\text{scorer}}$ are the runtimes of the model and scorer operating on data, $\mathcal{D}$. Table \ref{tab:BBbreakdown_recursive} asymptotically itemizes the algorithm. Truncation of the search space occurs with a probability $p$ of two recursions and $1-p$ of one recursion. Therefore, the recurrence relation is:
\[T(n) = p \cdot 2T\left(\frac{n}{2}\right) + (1-p) \cdot T\left(\frac{n}{2}\right) + O(T_{\text{model}} + T_{\text{scorer}})\]
The model and scorer operate on $\mathcal{D}$, so their runtimes do not scale with $n$ and thus can be represented as  $O(1)$ for each subproblem. This simplifies the equation to:
\begin{center}
$T(n) = (p + 1)T\left(\frac{n}{2}\right) + O(1)$
\end{center}
Applying the  Master Theorem \cite{cormen2022introduction} with $a= p + 1$ (number of recursive calls), $b=2$ (factor by which the problem size is divided) and $f(n) = O(1)$: 
$\log_b{a} = \log_2{(p + 1)}$
 For $\log_2{(p + 1)} < 0 $ to hold, $p$ would need to be less than $0$, probabilistically impossible. Therefore, we have $c < \log_2{(p + 1)} $, and case 1 of the Master Theorem applies:
\[
T(n) = \Theta(n^{\log_2{(p + 1)}})
\]

Thus, Binary Bleed's asymptotic runtime is $\Theta(n^{\log_2{(p + 1)}})$. 

\begin{figure}[ht]
\centering
\begin{minipage}[b]{0.48\columnwidth}
    \begin{table}[H]
    \centering
    \tiny 
    \begin{tabular}{| c | c | c |}
    \hline
    \textbf{Step} & \textbf{Line} & \textbf{Complexity} \\
    \hline
    Initialization &  1-2 & \( O(1) \) \\
    \hline
    Base Case &  3-4 & \( O(1) \) \\
    \hline
    Middle/Check &  5-7 & \( O(1) \) \\
    \hline
     Model/Score &  8 & \(  T_{\text{model}} + T_{\text{scorer}} \) \\
    \hline
    Update Min/Max  &  10-16 & \( O(1) \) \\
    \hline
    
    Bound Check & 17,20 & \( O(1) \) \\
    \hline
    Recursion &  18-23 & - \\
    \hline
    \end{tabular}
    \caption{Binary Bleed \( k \) Search primitive operations for asymptotic analysis.}
    \label{tab:BBbreakdown_recursive}
    \end{table}
\end{minipage}
\hfill
\centering
\begin{minipage}[b]{0.48\columnwidth}
    \begin{figure}[H]
    \centering
    \begin{tikzpicture}[
            level distance=4mm, 
            level 1/.style={sibling distance=22mm},
            level 2/.style={sibling distance=12mm}, 
            level 3/.style={sibling distance=8mm}, 
            every node/.style={circle, draw, scale=0.56}
    ]
    \node (a) {6}
      child {node (b) {3}
        child {node (c) {2}
          child {node (d) {1}}
          child[missing]
        }
        child {node (e) {5}
          child {node (f) {4}}
          child[missing]
        }
      }
      child {node (g) {9}
        child {node (h) {8}
          child {node (i) {7}}
          child[missing]
        }
        child {node (j) {11}
          child {node (k) {10}}
          child[missing]
        }
      };
    \foreach \x/\num in {a/0, b/1, c/2, d/3, e/4, f/5, g/6, h/7, i/8, j/9, k/10}
      \node at (\x) [circle, fill=red!30, scale=0.6, xshift=-7mm] {\num};
    \foreach \x/\num in {d/0, c/1, b/2, f/3, e/4, a/5, i/6, h/7, g/8, k/9, j/10}
      \node at (\x) [circle, fill=green!30, scale=0.6, yshift=7mm] {\num};
    \foreach \x/\num in {d/0, c/1, f/2, e/3, b/4, i/5, h/6, k/7, j/8, g/9, a/10}
      \node at (\x) [circle, fill=blue!30, scale=0.6, xshift=7mm] {\num};
    \end{tikzpicture}
    \caption{Traversal ordering sort of pre-order (red), in-order (green), and post-order (blue).}
    \label{fig:bst_traversal}
    \end{figure}
\end{minipage}
\end{figure}
  
\begin{table}[h]
\centering
\caption{Chunks Pre/Post Traversal Sorts on two resources  }
\label{tab:bst_traversal}
\scriptsize
\begin{tabular}{@{}ccc@{}}
\toprule
Order  & Operation 1                           & Operation 2 \\ \midrule

\multicolumn{1}{r}{\textbf{ T1}}  & \multicolumn{1}{c}{\cellcolor{blue!7}Traversal Order Sort} & \multicolumn{1}{c}{\cellcolor{yellow!7}Chunk Ks by Resource Count} \\
\cmidrule(lr){2-2} \cmidrule(lr){3-3}
In    &  1, 2, 3, 4, 5, 6, 7, 8, 9, 10, 11   & [1, 2, 3, 4, 5, 6] [7, 8, 9, 10, 11]   \\
\cellcolor{red!5}Pre   & \cellcolor{red!5}6, 3, 2, 1, 5, 4, 9, 8, 7, 11, 10    & [6, 3, 2, 1, 5, 4] [9, 8, 7, 11, 10]   \\
Post  & 1, 2, 4, 5, 3, 7, 8, 10, 11, 9, 6    & [1, 2, 4, 5, 3, 7] [8, 10, 11, 9, 6]   \\ 
\cmidrule(lr){2-2} \cmidrule(lr){3-3}

\multicolumn{1}{r}{\textbf{ T2}}  & \multicolumn{1}{c}{\cellcolor{blue!7}Traversal Order Sort} & \multicolumn{1}{c}{\cellcolor{green!7}Chunk Ks by Alg. \ref{alg:chunk_k}} \\
\cmidrule(lr){2-2} \cmidrule(lr){3-3}
In    & 1, 2, 3, 4, 5, 6, 7, 8, 9, 10, 11   & [1, 3, 5, 7, 9, 11] [2, 4, 6, 8, 10]   \\
\cellcolor{red!5}Pre   &\cellcolor{red!5} 6, 3, 2, 1, 5, 4, 9, 8, 7, 11, 10     & [3, 1, 5, 9, 7, 11]  [6, 2, 4, 8, 10]   \\
Post  & 1, 2, 4, 5, 3, 7, 8, 10, 11, 9, 6    & [1, 5, 3, 7, 11, 9]  [2, 4, 8, 10, 6]   \\ 
\midrule  
\multicolumn{1}{r}{\textbf{T3}} & \multicolumn{1}{c}{\cellcolor{yellow!7} Chunk Ks by Resource Count} & \multicolumn{1}{c}{\cellcolor{blue!7}Traversal  Order Sort} \\
\cmidrule(lr){2-2} \cmidrule(lr){3-3}
In    & [1, 2, 3, 4, 5, 6] [7, 8, 9, 10, 11]   & [1, 2, 3, 4, 5, 6] [7, 8, 9, 10, 11]   \\
\cellcolor{red!5}Pre   &  [1, 2, 3, 4, 5, 6] [7, 8, 9, 10, 11]     & \cellcolor{red!5}[4, 2, 1, 3, 6, 5]  [9, 8, 7, 11, 10]   \\
Post  &  [1, 2, 3, 4, 5, 6] [7, 8, 9, 10, 11]     & [1, 3, 2, 5, 6, 4] [7, 8, 10, 11, 9] \\
\cmidrule(lr){2-2} \cmidrule(lr){3-3}

\multicolumn{1}{r}{\textbf{T4}} & \multicolumn{1}{c}{\cellcolor{green!7} Chunk Ks by Alg. \ref{alg:chunk_k}} & \multicolumn{1}{c}{\cellcolor{blue!7}Traversal  Order Sort} \\
\cmidrule(lr){2-2} \cmidrule(lr){3-3}
In    & [1, 3, 5, 7, 9, 11] [2, 4, 6, 8, 10]   & [1, 3, 5, 7, 9, 11] [2, 4, 6, 8, 10]   \\
\cellcolor{red!5}Pre   &  [1, 3, 5, 7, 9, 11] [2, 4, 6, 8, 10]    & \cellcolor{red!5}[7, 3, 1, 5, 11, 9] [6, 4, 2, 10, 8]   \\
Post  &  [1, 3, 5, 7, 9, 11] [2, 4, 6, 8, 10]    & [1, 5, 3, 9, 11, 7] [2, 4, 9, 10, 6]   \\ \bottomrule
\end{tabular}
\end{table}

\begin{algorithm}[t!]
\caption{Chunk $k$ values by Skip Mod Resource Count}\label{alg:chunk_k}
\begin{algorithmic}[1]
\Require \( \mathcal{K} \) (list of $k$), $num\_resources$ (resource count to split \( k \))
\Ensure List of chunks where each chunk contains integers assigned to a resource
\Function{ChunkKs}{$K, num\_resources$}
    \State $K\_chunks \gets []$
    \For{$i = 0$ \textbf{to} $num\_resources - 1$}
        \State $K\_chunks[i] \gets$ empty list
    \EndFor
    \For{$k = 0$ \textbf{to} $Ks - 1$}
        \State $resource\_id \gets k \mod num\_resources$
        \State Append $k$ to $K\_chunks[resource\_id]$
    \EndFor
    \State \textbf{return} $K\_chunks$
\EndFunction
\end{algorithmic}
\end{algorithm}

\subsection{ Binary Bleed Multi-threaded, Multi-rank}

A parallel implementation of $Binary\ Bleed$ can be obtained by extending the serial implementation depicted in Algorithm \ref{alg:binary_search_optimal_k} with the following changes: Make $k_{min}$, $k_{max}$ and the list of visited $k$ $rank_{seen}$ global using a distributed cache such as reddis, and $i_{left}$ and $i_{right}$ are now computed by each MPI rank or thread as $i_{left}=k_{min}+idx \times (k_{max}-k_{min})/size$ and $i_{rigth}=min(i_{left}+size, k_{max})$ where $idx$ is the thread index or MPI rank, and $size$ is the total number of threads or MPI ranks. The resulting parallel implementation will work well, but will not be effective when the list of $k$ is sparse.

A more robust algorithm can be obtained by replacing recursions in Algorithm \ref{alg:binary_search_optimal_k} by a $k$-sort, illustrated in Figure \ref{fig:bst_traversal}, where $k$ values are sorted using in-order, pre-order, or post-order binary tree traversal, indicated by the color on three sides of each rank's $k$ value.

Further, \( \mathcal{K} \) must be chunked into available resources using Algorithm \ref{alg:chunk_k}.  By appropriately sorting and chunking the $k$ values, the Binary Bleed algorithm can efficiently operate in a multi-threaded and multi-rank environment, enhancing its scalability and performance.

\textbf{ Logistics of Chunking \( \mathcal{K} \) and Traversal Sort }: To illustrate the optimal function sequence, assume $k=[1-10]$, with two operating resources. The four splits of the data are:

\begin{enumerate}
  \item Sort, then data chunk by resources (Table \ref{tab:bst_traversal}'s \textbf{T1})
  \item Sort, then data chunk by Algorithm \ref{alg:chunk_k} (  Table \ref{tab:bst_traversal}'s \textbf{T2})
  \item Data chunk by resources, then sort (Table \ref{tab:bst_traversal}'s \textbf{T3})
  \item Data chunk by Algorithm \ref{alg:chunk_k}, then sort (Table \ref{tab:bst_traversal}'s \textbf{T4})
\end{enumerate}

Table  \ref{tab:bst_traversal} distributes $k$ on two resources, showing in-order traversal monotonically increases, leading to inadequate ordering for early termination of $k$.
Truncation removes unvisited smaller values, which is impossible for in-order traversal. Furthermore, \textbf{T1} demonstrates the insufficient distribution of \( k \) values across resources. For example, the second resource may find an optimal result in truncation for all \( k \) in the first resource, leading to one resource being idle while the other operates.
The remaining splits present \textbf{T3} as the least optimal, as the values are similarly partitioned to \textbf{T1}, resulting in idle higher optimal resources. \textbf{T2} and \textbf{T4} are distinguished by order, so operation logistics are a lower priority than the use of Algorithm \ref{alg:chunk_k} as a load-balanced partition. Table \ref{tab:bst_traversal} shows traversal selections are ambiguous. However, Section \ref{sec:results} indicates pre-order traversals visit fewer \( k \) values overall. Algorithm \ref{alg:threading} coordinates with Algorithm \ref{alg:chunk_k} to distribute and process $k$ values across multiple resources efficiently.

 \begin{algorithm}[h]
\caption{Multiple Ranks and Threads Binary Bleed}\label{alg:threading}
\begin{algorithmic}[1]
\Require $Ks$ (sorted $k$ values), $N$ (ranks in network), $t$ (threads per rank)
\State Initialize all ranks with unique identifiers and communication between ranks

\Procedure{InitializeRankKs}{$Ks, N, t, data$}

    \State $K_{\text{chunks}} \gets$ \Call{ChunkKs}{$Ks, N$}
 
    \For{each rank $n$ in the network}
        \State Initialize $Ks_{\text{bst}} \gets []$ 
        
        \State \Call{TraversalSort}{$K_{\text{chunks}}[n], length(Ks), \&Ks_{\text{bst}}$}
        \State \Call{StartThreads}{$data, Ks_{\text{bst}}, t$}  
     
    \EndFor
\EndProcedure
\Procedure{StartThreads}{ $data, Ks, num_{\text{threads}} $}  
    \State $ Ks\_size \gets |Ks|$, $mutex \gets \text{CreateMutex()}$ 
    
    \For{$i \gets 0 \text{ to } Ks\_size$ } 
        \State \textbf{start thread} \Call{BinaryBleedMulti}{ 
        
        $Ks_{\text{bst}}[i \% num_{\text{threads}}],  data, \&k_{\text{optimal}}, mutex$
        }
    \EndFor
    \State \textbf{wait for all threads to finish, then} \textbf{process} $k_{\text{optimal}}$
\EndProcedure
\Procedure{BroadcastK}{$k_{\text{optimal}}, sender$}
    \For{each rank $n$ in the network}
        \If{$n \neq sender$}  Send $k_{\text{optimal}}$ to rank $n$
        \EndIf
    \EndFor
\EndProcedure
\Procedure{ReceiveKCheck}{ $k_{\text{receiver}}, receiver$}
    \If{message from network}  Receive $k_{\text{sender}}, sender$
    \Else \textbf{ return} $ k_{\text{receiver}}$
    \EndIf
    
    \If{$ k_{\text{receiver}} < k_{\text{sender}} $}
       $ k_{\text{receiver}} \gets k_{\text{sender}}$
    \Else \textbf{ Send} $k_{\text{receiver}}$ to rank $sender$
    \EndIf
\EndProcedure
\end{algorithmic}
\end{algorithm}

First, in Algorithm \ref{alg:threading}, the  $k$ list, number of ranks, and threads per rank are needed. Ranks are initialized with IDs and communication setup (lines 1-2). Algorithm  \ref{alg:chunk_k} chunks the  $k$ (line 3). Each rank initializes a binary search tree (BST) (lines 4-5) of sorted  $k$ and starts threads on the data (lines 7-9). The thread initializes variables for $|\mathcal{K}|$ and a mutex (lines 10-11). It iterates over the threads to start Binary Bleed on each $k$ value, waits for completion, and processes the optimal value (lines 12-16). The broadcast function iterates the ranks, sending the optimal  $k$ to others (lines 17-22). The function for receiving communication is defined (line 23). On receiving a message, the rank checks if the received  $k$ exceeds the current optimal, updating or returning the higher optimal as needed (lines 24-30). The second change needed for parallel Binary Bleed is the communication of pruned $k$ values to other resources. Algorithm \ref{alg:threading}, in conjunction with Algorithm \ref{alg:binary_search_multi}, contains the set of calls to complete the communication of pruning in parallel and distributed Binary Bleed.

\begin{algorithm}[t!]
\caption{Binary Bleed $k$ Search, Multi-rank \& Thread }\label{alg:binary_search_multi}
\begin{algorithmic}[1]
\Require $k$ (value to compute), $data$ (dataset ), $model$ (calculator), $scorer$ (quantifies output), $T_{\text{select\_k}}$ (minimum $k$ score), $rank\_id$ (id of rank)
\Ensure Optimal $k$ value and its prediction score
\State $k_{\text{optimal}} \gets  null$  

\Function{BinaryBleedMulti}{
$k, k_{\text{optimal}}, data, mutex, T_{\text{select\_k}}$}
    \State Initialize $model, scorer$, $report \gets 0$  
    \State $\text{Lock}(mutex)$ \textbf{and}  $k_{\text{received}} \gets k_{\text{optimal}}$  \textbf{then}   $\text{Unlock}(mutex)$

    \State $k_{\text{received}} \gets$ \Call{ReceiveKCheck}{    
    $k_{\text{receiver}}, rank\_id$ }
    
    \State $\text{Lock}(mutex)$
        \If{$k_{\text{received}} \neq k_{\text{optimal}}$}
            \If{$\neg k_{\text{optimal}}$ \textbf{or} $k_{\text{received}} > k_{\text{optimal}}$} $k_{\text{optimal}} \gets k_{\text{received}}$
            \Else  \textbf{ } $report \gets 1$
            \EndIf
        \EndIf
    \State $\text{Unlock}(mutex)$

    \If{$report$} \Call{BroadcastK}{$k_{\text{optimal}}, rank\_id$} \textbf{and} $report \gets 0$ 
    \Else
        \If{$k_{\text{optimal}} > $k$ $ } \textbf{return} \EndIf
    \EndIf
    \State $score \gets scorer(model(data, k))$
    \If{$score \geq k_{\text{select\_threshold}}$}  $\text{Lock}(mutex)$
         \If{$\neg k_{\text{optimal}}$ \textbf{or} $k > k_{\text{optimal}}$} $k_{\text{optimal}} \gets k$ \textbf{and}  $report \gets 1$
        \EndIf
        \State $\text{Unlock}(mutex)$
        \If{$report$} \Call{BroadcastK}{$k_{\text{optimal}}, rank\_id$}
        \EndIf
    \EndIf
\EndFunction
\end{algorithmic}
\end{algorithm}

Algorithm \ref{alg:binary_search_multi} requires the \( k \) range, data, model, scorer, selection threshold, and rank ID. Lines (1-3) initialize the optimal \( k \), define the Binary Bleed multi-function, and initialize the model, scorer, and report flag. Line (4) locks the mutex, sets the received \( k \) as optimal, and unlocks the mutex. Line (5) checks network messages and sets the received \( k \). Lines (6-12) lock the mutex, compare optimal and received \( k \), update the optimal if different, and possibly set the report flag before unlocking the mutex. Lines (13-17) check if a report is needed for other resources and reset the report flag if necessary. Line (18) operates the model, data, and scorer on \( k \). Lines (19-26) lock the mutex if the score exceeds the threshold, update the optimal \( k \) if larger, and broadcast if the optimal was updated. In multi-threading, the rank propagates the optimal \( k \) from the finding thread to others, pruning smaller \( k \) values. In multi-rank, any rank propagates the optimal \( k \) to other ranks. In both cases, threads report to ranks, which then report to other ranks and threads. In an HPC system, resources are processes or compute ranks. In multi-threading, the optimal  \( k \) a thread finds is propagated by the controlling rank to all threads, which prune smaller  \( k \) values. In multi-rank, any rank finding the optimal  \( k \) shares it with all ranks. Threads report to ranks, then communicate the optimal  \( k \) to other ranks and threads.

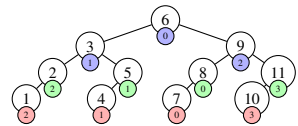
\begin{figure}[h]
\centering
\begin{minipage}[b]{0.48\columnwidth}
    \begin{minipage}{\columnwidth}
    \begin{figure}[H]
    \centering
    \begin{tikzpicture}[
        level distance=3.5mm, 
        level 1/.style={sibling distance=20mm},
        level 2/.style={sibling distance=10mm}, 
        level 3/.style={sibling distance=7mm}, 
        every node/.style={circle, draw, scale=0.6}
    ]
    \node (a) {6}
      child {node (b) {3}
        child {node (c) {2}
          child {node (d) {1}}
          child[missing]
        }
        child {node (e) {5}
          child {node (f) {4}}
          child[missing]
        }
      }
      child {node (g) {9}
        child {node (h) {8}
          child {node (i) {7}}
          child[missing]
        }
        child {node (j) {11}
          child {node (k) {10}}
          child[missing]
        }
      };
    \foreach \x/\num in {i/0, f/1, d/2, k/3}
      \node at (\x) [circle, fill=red!30, scale=0.6, yshift=-6mm] {\num};
    \foreach \x/\num in {h/0, e/1, c/2,  j/3}
      \node at (\x) [circle, fill=green!30, scale=0.6, yshift=-6mm] {\num};
    \foreach \x/\num in { a/0, b/1, g/2}
      \node at (\x) [circle, fill=blue!30, scale=0.6, yshift=-6mm] {\num};
    \end{tikzpicture}
    \caption{Vanilla: Part 1}
    \label{fig:bst1}
    \end{figure}
    \end{minipage}%
\end{minipage}
\hfill
\centering
\begin{minipage}[b]{0.48\columnwidth}
    \begin{minipage}{\columnwidth}
    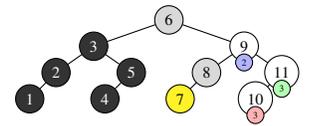
\begin{figure}[H]
    \centering
    \begin{tikzpicture}[
        level distance=3.5mm, 
        level 1/.style={sibling distance=20mm},
        level 2/.style={sibling distance=10mm}, 
        level 3/.style={sibling distance=7mm}, 
        every node/.style={circle, draw, scale=0.6}
    ]
    \node [fill=gray!30] (a) {6}
      child {node[fill=black!80, text=white] (b) {3}
        child {node[fill=black!80, text=white] (c) {2}
          child {node[fill=black!80, text=white] (d) {1}}
          child[missing]
        }
        child {node[fill=black!80, text=white] (e) {5}
          child {node[fill=black!80, text=white] (f) {4}}
          child[missing]
        }
      }
      child {node (g) {9}
        child {node[fill=gray!30]  (h) {8}
          child {node[fill=yellow!90] (i) {7}}
          child[missing]
        }
        child {node (j) {11}
          child {node (k) {10}}
          child[missing]
        }
      };
    \foreach \x/\num in { k/3}
      \node at (\x) [circle, fill=red!30, scale=0.6, yshift=-6mm] {\num};
    \foreach \x/\num in {   j/3}
      \node at (\x) [circle, fill=green!30, scale=0.6, yshift=-6mm] {\num};
    \foreach \x/\num in {  g/2}
      \node at (\x) [circle, fill=blue!30, scale=0.6, yshift=-6mm] {\num};
    \end{tikzpicture}
    \caption{ Vanilla: Part 2}
    \label{fig:bst2}
    \end{figure}
    \end{minipage}
\end{minipage}
\end{figure}

\begin{wrapfigure}{l}{0.53\columnwidth}
    \centering
    \includegraphics[width=0.56\columnwidth]{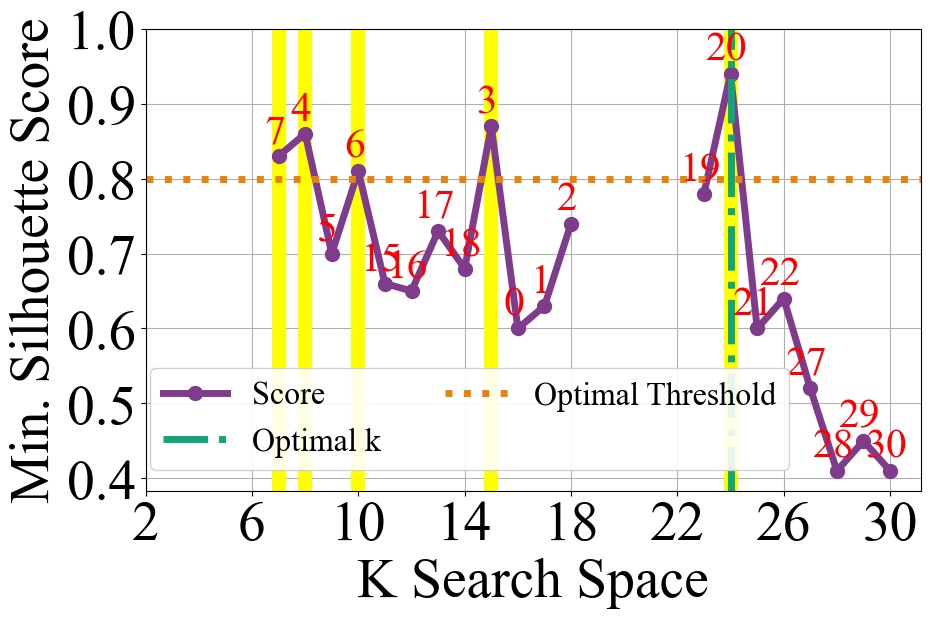}
    \caption{ Binary Bleed Vanilla.}
    \label{fig:multi}
\end{wrapfigure}

\textbf{Operation Dynamics}: Figure \ref{fig:bst1} Pre-order sorts \( k =  \{1,2,\cdots,11\}\) on three resources after Algorithm \ref{alg:chunk_k} (\textbf{T4} \ref{tab:bst_traversal}). $\forall k$ are the possible optimal.  Resource allocation (r,g,b) is by visit order.  In part 2, Figure \ref{fig:bst2}, 
the score of $k=7$ is greater than the threshold, so it is optimal,  and the scores of $k=\{6,8\}$ are less than the threshold.  $k=\{1,2,\cdots,5\}$ are pruned for being less than the optimal. The upper $k$ range, $k=\{9,10,11\}$ continues. Figure \ref{fig:multi} schedules \( k \) values are chunked by four resources, then pre-order sorted. 
The graph shows the optimal selection threshold as a dashed gold line.  Yellow vertical bars indicate truncation is reported to other resources, where a \( k \) score passed the threshold. The gold threshold is crossed at four values: \( k = \{7, 8, 10, 24\} \). When the first \( k = 8 \) passes the threshold, all \( k < 8 \) are pruned. However, the implementation shown does not prune \( k \) values after the model begins execution; therefore, \( k = 7 \) is complete. The \( k \) values [7, 10] have no remaining \( k \) values to prune. Due to pre-order sorting, \( K = 24 \) is run before \( k =  \{18,19,\cdots,22\}\). Since \( k = 24 \) is above the selection threshold, those lower priority \( k \) values are pruned. The upper \( k \) range continues without crossing the selection threshold, so \( K = 24 \) remains optimal.

\subsection{ Binary Bleed Early Stop}
\label{Early_stopping}
Specific domains allow for an additional heuristic. Rather than pruning the lower \( k \) values, setting a bound on the other extreme of the data will allow for truncation of the upper values. This heuristic is based on domain knowledge, where if any scores cross the stop threshold, they will never rise to cross the selection threshold, and subsequent \( k \) values can be ignored. Mathematically, it can be represented as:
\[
k_{\text{optimal}} = \max \left\{ k \in \{1, 2, \ldots, K\} : \forall i \leq k, S(k_i) > U \right\}
\]
where \( S(k_i) \) is the score of the \( i^{th} \) \( k \), and \( U \) is the stop bound.

\begin{figure}[h]
\begin{minipage}[b]{0.48\columnwidth}
    \begin{minipage}{\columnwidth}
    \begin{figure}[H]
    \centering
    \begin{tikzpicture}[
        level distance=3.5mm, 
        level 1/.style={sibling distance=20mm},
        level 2/.style={sibling distance=10mm}, 
        level 3/.style={sibling distance=7mm}, 
        every node/.style={circle, draw, scale=0.6}
    ]
    \node (a) {6}
      child {node (b) {3}
        child {node (c) {2}
          child {node (d) {1}}
          child[missing]
        }
        child {node (e) {5}
          child {node (f) {4}}
          child[missing]
        }
      }
      child {node (g) {9}
        child {node (h) {8}
          child {node (i) {7}}
          child[missing]
        }
        child {node (j) {11}
          child {node (k) {10}}
          child[missing]
        }
      };
    \foreach \x/\num in {e/0, d/1, g/2}
      \node at (\x) [circle, fill=red!30, scale=0.6, yshift=-6mm] {\num};
    \foreach \x/\num in {a/0, c/1, k/2}
      \node at (\x) [circle, fill=green!30, scale=0.6, yshift=-6mm] {\num};
    \foreach \x/\num in { i/0, b/1, j/2}
      \node at (\x) [circle, fill=blue!30, scale=0.6, yshift=-6mm] {\num};
      \foreach \x/\num in { h/0, f/1}
      \node at (\x) [circle, fill=orange!30, scale=0.6, yshift=-6mm] {\num};
      
    \end{tikzpicture}
    \caption{Early Stop: Part 1}
    \label{fig:bst3}
    \end{figure}
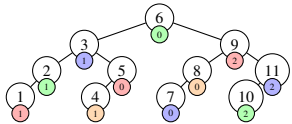
    \end{minipage}%
\end{minipage}
\hfill
\centering
\begin{minipage}[b]{0.48\columnwidth}
    \begin{minipage}{\columnwidth}
    \begin{figure}[H]
    \centering
    \begin{tikzpicture}[
        level distance=3.5mm, 
        level 1/.style={sibling distance=20mm},
        level 2/.style={sibling distance=10mm}, 
        level 3/.style={sibling distance=7mm}, 
        every node/.style={circle, draw, scale=0.6}
    ]
    \node [fill=gray!30] (a) {6}
      child {node[fill=black!80, text=white]  (b) {3}
        child {node[fill=black!80, text=white]  (c) {2}
          child {node[fill=black!80, text=white]  (d) {1}}
          child[missing]
        }
        child {node[fill=yellow!90]  (e) {5}
          child {node[fill=black!80, text=white] (f) {4}}
          child[missing]
        }
      }
      child {node[fill=black!80, text=white] (g) {9}
        child {node[fill=purple!70, text=white]  (h) {8}
          child {node[fill=gray!30] (i) {7}}
          child[missing]
        }
        child {node[fill=black!80, text=white] (j) {11}
          child {node[fill=black!80, text=white] (k) {10}}
          child[missing]
        }
      };

    \end{tikzpicture}
    \caption{Early Stop: Part 2. }
    \label{fig:bst4}
    \end{figure}
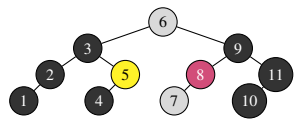
    \end{minipage}%
\end{minipage}%
\end{figure}

 Figure \ref{fig:bst3} Pre-order sorts $k=[1-11]$ after Algorithm \ref{alg:chunk_k} (\textbf{T4} \ref{tab:bst_traversal}), on four resources. In part 2, Figure \ref{fig:bst4}, $k=5$ exceeds the selection threshold, so the optimal is set and $k=[1-4]$ (dark nodes) are pruned.  $k=8$ crosses the stop threshold, so $k=[9-11]$ are pruned. Therefore, the optimal remains 5.

\subsection{ Additional Considerations }

In multi-resource computing, an optimal $k$ may be larger than an executing \( k \). For long computations, checks can be pushed into the model to terminate such $k$ early.
Consider the scoring distribution, like silhouettes, which can speed up the process. True optimal values are found faster when scores above the threshold resemble a square wave. Mathematically:

\[
S(k_i) = \frac{\text{sgn}(k_0 - k_i) + 1}{2}
\]

where \( S(k_i) \) is the evaluation score at the \( i^{th} \) index of \( k \) values, \( k_0 \) is the optimal \( k \), and \(\text{sgn}\) is the signum function, which evaluates to \( +1 \) for \( k_i < k_0 \) and \( -1 \) for \( k_i \geq k_0 \).
In the worst case, a Laplacian score distribution will peak at the optimal \( k \) while other \( k \) scores will be below the score selection threshold. Lower \( k \) values may be pruned if the peak is visited before lower values. Otherwise, all values will be visited in the order of the sort. Despite the score distribution, Binary Bleed will not visit more \( k \) values than a linear search.


\section{Experimental Results }
\label{sec:results}
We gauge the efficacy of unsupervised learning methods in single-node and distributed settings. We evaluate  NMF and K-means in the former, then NMF and RESCAL in the latter.
\begin{figure}[h]
    \centering
    \begin{minipage}[b]{0.48\columnwidth}
        \centering
        \includegraphics[width=\columnwidth]{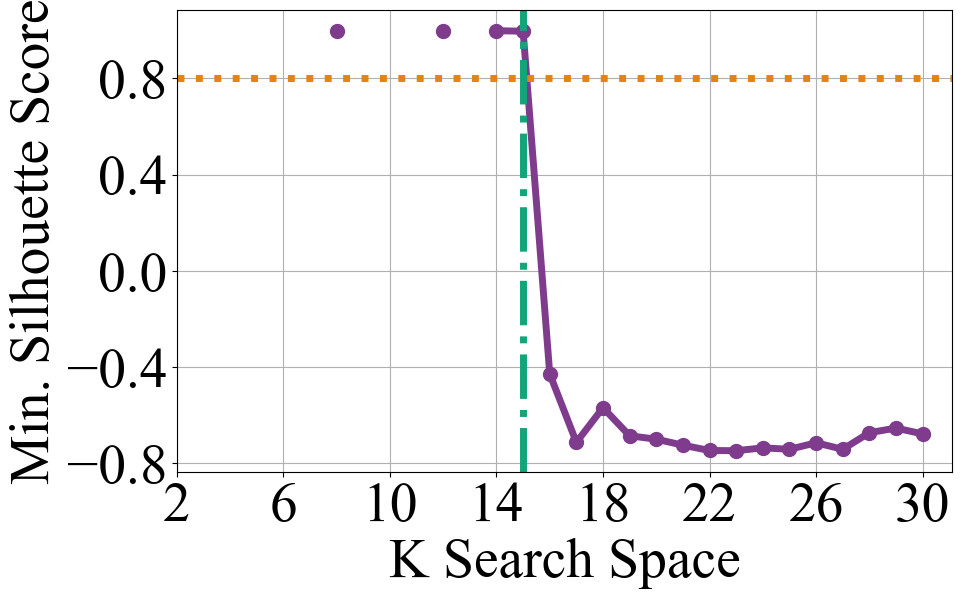}
    \end{minipage}
    \hfill
    \begin{minipage}[b]{0.48\columnwidth}
        \centering
        \includegraphics[width=\textwidth]{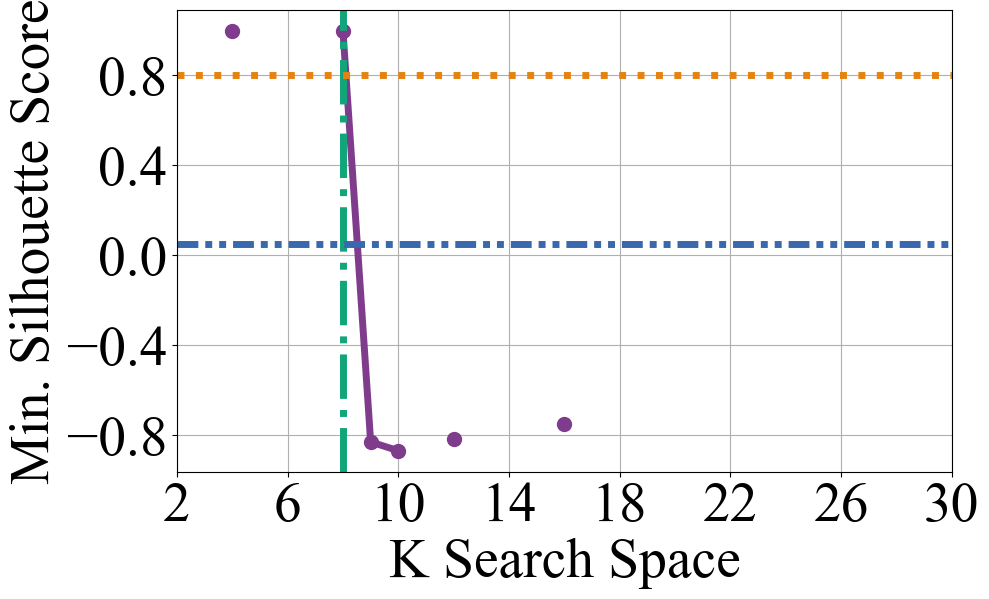}
    \end{minipage}
    \vskip\baselineskip
    \begin{minipage}[b]{0.48\columnwidth}
        \centering
        \includegraphics[width=\columnwidth]{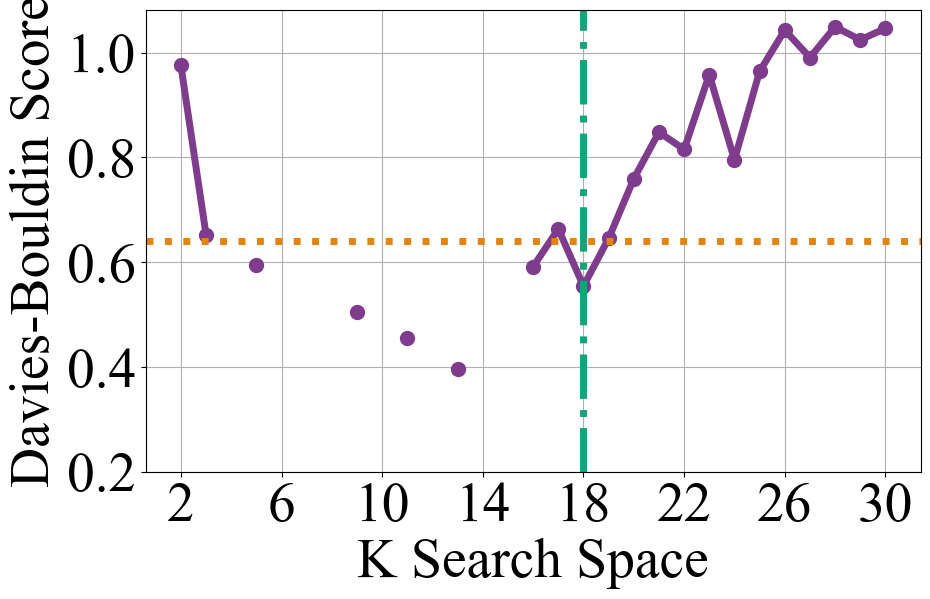}
    \end{minipage}
    \hfill
    \begin{minipage}[b]{0.48\columnwidth}
        \centering
        \includegraphics[width=\columnwidth]{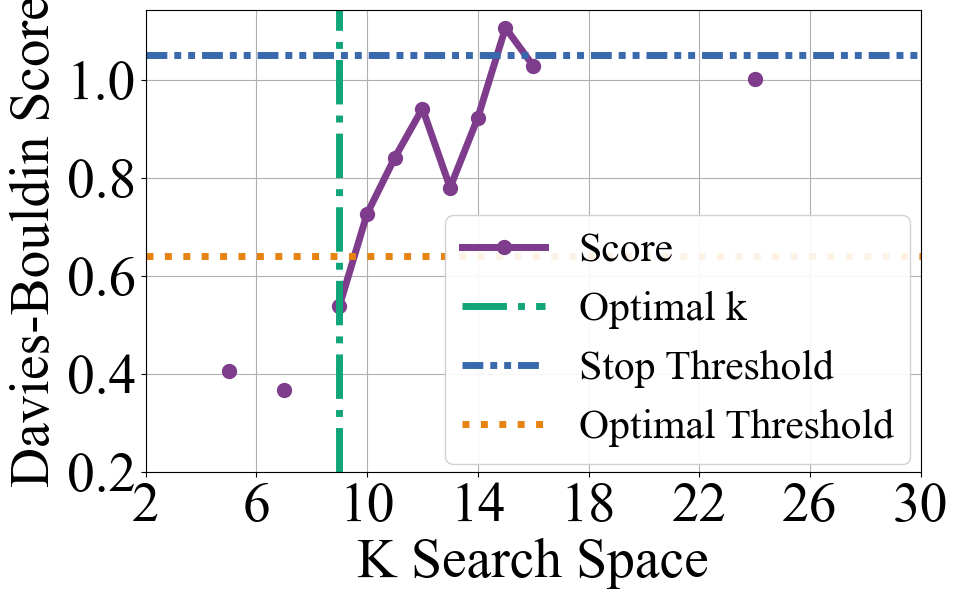}
    \end{minipage}
    \caption{NMFk (top row) and K-means (bottom row), Vanilla (Left) and Early stopping (Right).  $\forall k_{optimal}=k_{true}$.}
    \label{fig:2x2grid}
\end{figure}

\subsection{Single-node Setting}
\begin{figure}[h]
    \centering
    \begin{minipage}[b]{0.49\columnwidth}
         \begin{figure}[H]
        \centering
         \includegraphics[width=\columnwidth]{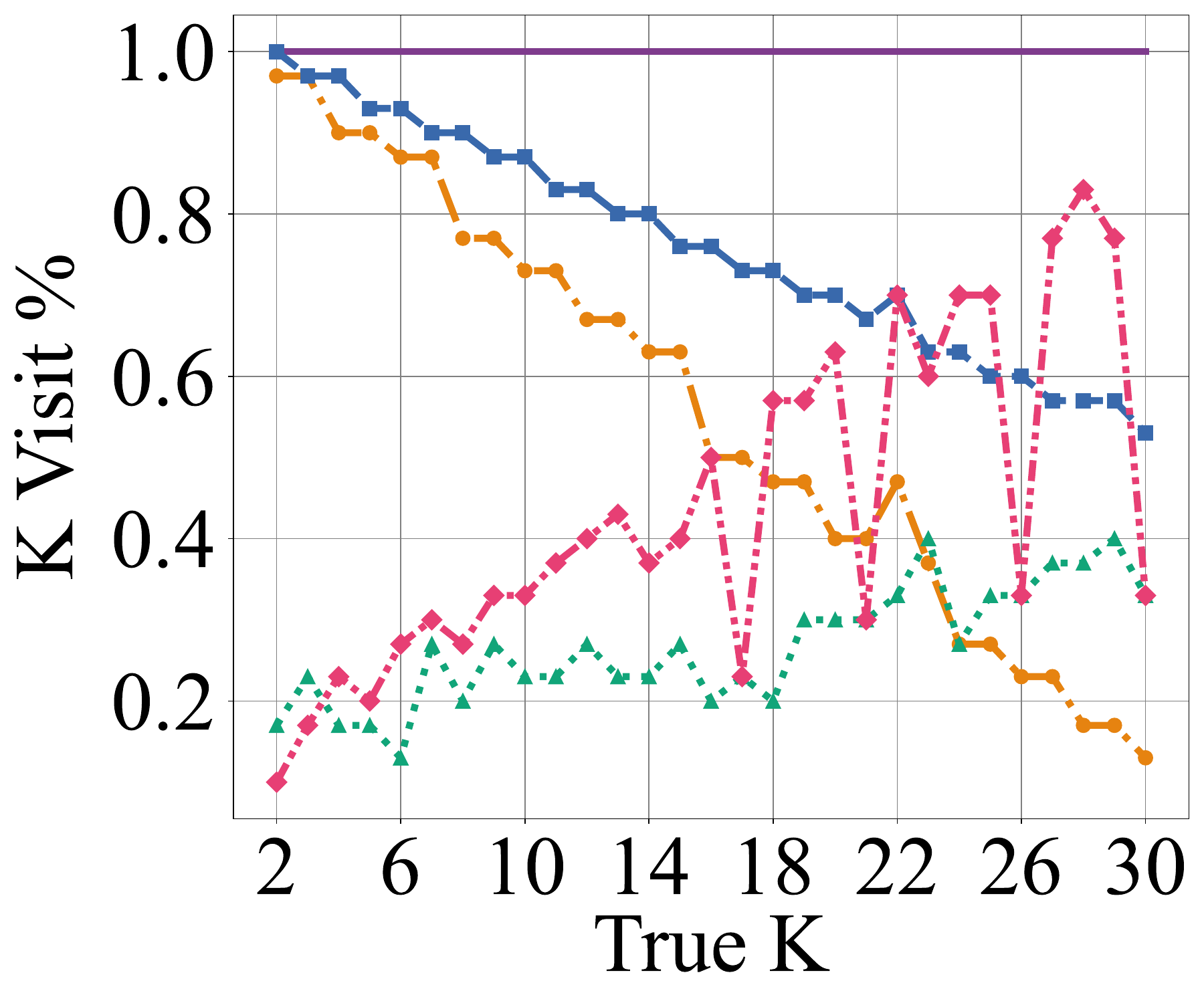}
        
            \label{fig:preorder_results}
        \end{figure}
    \end{minipage}
\hfill
\centering
    \begin{minipage}[b]{0.49\columnwidth}
          \begin{figure}[H]
            \centering
             \includegraphics[width=\columnwidth]{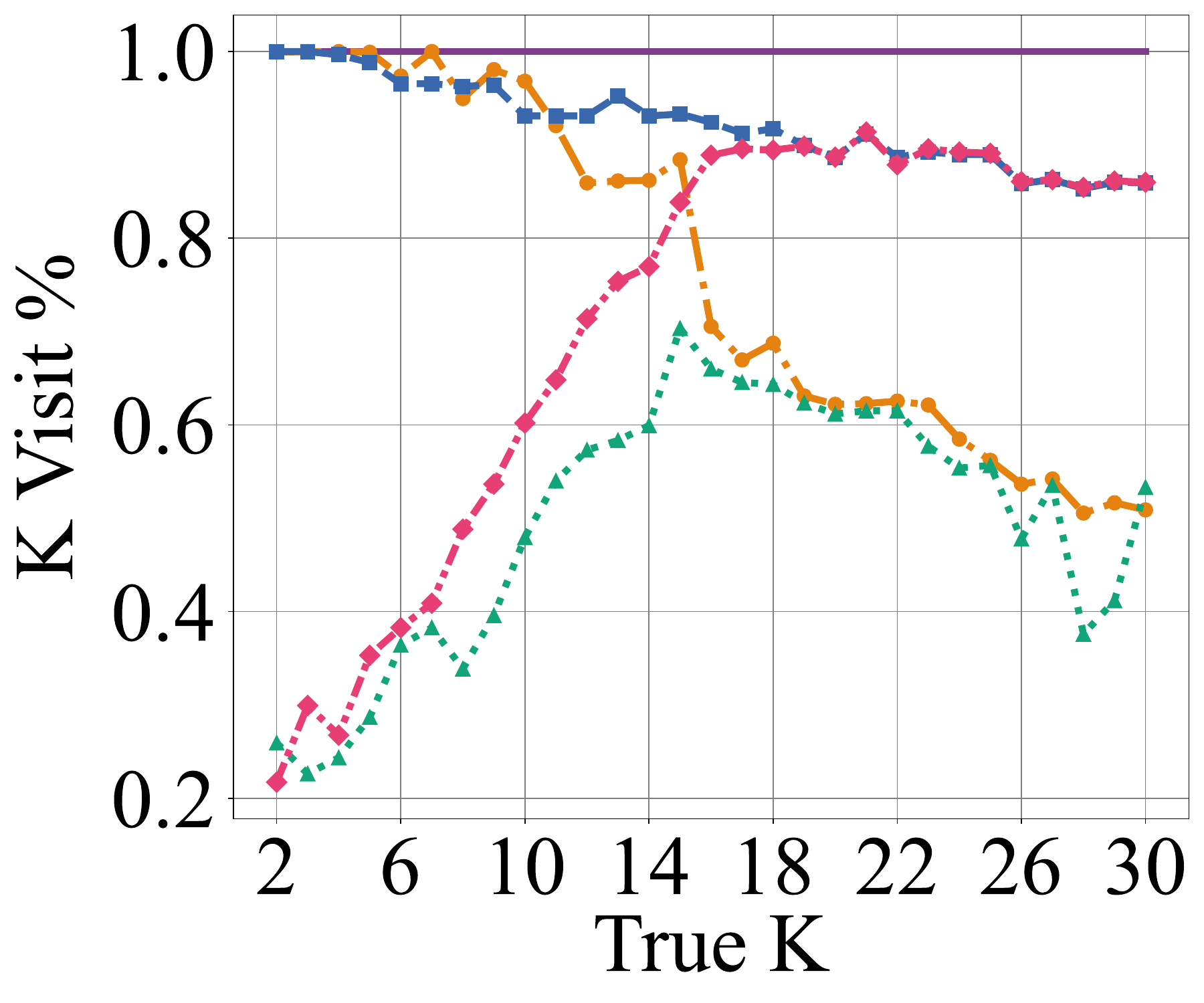}
            
                \label{fig:kmeans_ressults}
            \end{figure}
    \end{minipage}
\hfill
\vspace{0pt} 
\centering
    \begin{minipage}[b]{0.97\columnwidth}
          \begin{figure}[H]
            \centering
             \includegraphics[width=\columnwidth]{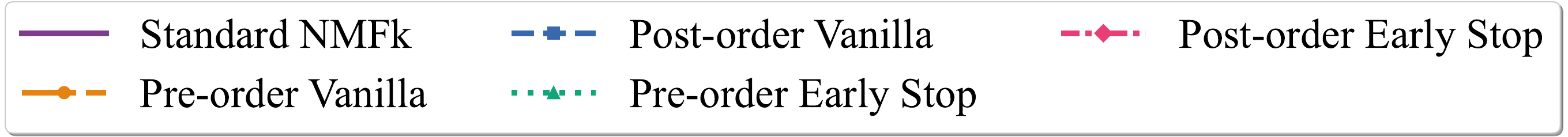}
            
                \label{fig:kmeans_ressults}
            \end{figure}
    \end{minipage}
    
\caption{ Standard NMFk (left) and K-means (right), Vanilla, and Early Stop over Pre-order, and Post-order traversal sorting.}
\label{fig:combined_results_kmean_nmfk}
\end{figure}

\textbf{NMFk}: Data was generated using a synthetic data generator with random Gaussian features for a predetermined $k$, where $k_{true}=\{2,3,\cdots,30\}$.  The shape of 30 matrices was sized 1000 by 1100, resulting in 1.1 million non-negative entries. The $k_{true}$ predetermined the number of clusters in the data, typically showing a drastic drop-off in the scoring metric for subsequent $k$ values. The evaluation criterion was the silhouette scores of the proposed clusters for the visited $k$. For the NMFk Binary Bleed trials, three separate instances were operated over the same data: Standard NMFk, Binary Bleed Vanilla, and Binary Bleed Early Stopping. Each instance was evaluated on all 30 matrices with $k$ search space $\mathcal{K}=\{2,3,\cdots,30\}$. The results corresponding to NMFk Vanilla and NMFk Early Stopping for $k = 15$ and $k = 8$, respectively, are Figure~\ref{fig:2x2grid}'s top row. It can be seen that Binary Bleed pruned multiple $k$ values, whereas the standard method must visit all $\mathcal{K}$.
In the overview, Figure \ref{fig:combined_results_kmean_nmfk}, orange and blue lines show the Binary Bleed Vanilla algorithm and the downward trend of $k$ visits relative to the standard, where Pre-order finds the optimal $k$ in fewer overall visits. Similarly, Early Stop in pink and green have lower overall visits than Vanilla, with Pre-order benefiting more despite a slightly increasing trend for both Early Stop lines over $k_{true}$. Interestingly, the post-order Early Stop between being as fast as the Pre-order and visiting more $k$ than the Binary Bleed Vanilla, which can be attributed to the number of compute resources paired with which values are $k_{true}$ and the ordering of $\mathcal{K}$.  Overall, the algorithms visit the following percentages of $\mathcal{K}$: Pre-order Vanilla: 56\%, Post-order Vanilla: 76\%, Pre-order Early Stop: 27\%, Post-order Early Stop: 44\%, while Standard NMFk visits 100\% of $\mathcal{K}$.  This shows Pre-ordering $\mathcal{K}$ with Early Stop executes fastest.

\textbf{K-means}: The data was generated by simulating Gaussian-distributed clusters with a standard deviation of .5 and the true $k$ cluster count. Overlaid random noise introduces variability and ensures robustness. $k_{true}=\{2,3,\cdots,30\}$ and  $\mathcal{K}=\{2,3,\cdots,30\}$. Given the stochastic scoring, we cluster fifty times for each $k_{true}$ on every method-ordering pair (Vanilla, K-means Standard, Early Stop) in Pre-order and Post-order configurations. Davies-Bouldin scoring is used to evaluate the cluster quality and to determine the correct $k$. The Root Mean Square Error (RMSE) scores of correct $k$ were as follows: Post-Order Early Stop: 1.08, Pre-order Early Stop: 2.11, Post-Order Vanilla: 1.08, Pre-order Vanilla: 1.72, and baseline Standard k-means: 1.32. These results indicate the accuracy in identifying $k_{true}$, with all Binary Bleed iterations having RMSEs of 0.79 or less, close to the standard RMSE of 1.32.
 
 K-means Vanilla and K-means Early Stopping results for $k = 18$ and $k = 9$, respectively, are shown in Figure~\ref{fig:2x2grid}. In the overview, Figure \ref{fig:combined_results_kmean_nmfk}, Early Stop, pink and green, dominate the speed-up in lower $k$, but the dominating factor transitions to Pre-order sorting after $k=15$. Average $k$ visit percentages are-- Pre-order Vanilla: 77\%, Post-order Vanilla: 92\%, Pre-order Early Stop: 50\%, Post-order Early Stop: 71\%. As reported, the percent $k$ visited shows the reduced amounts $k$ visits needed to complete the optimization process.

\subsection{Multi-node Setting}
 We demonstrate a reduction in $k$ visits for topic modeling of over 2 million scientific abstracts from arXiv with NMFk from \cite{barron2024cybersecurity}. We used the LANL Chicoma super-computer cluster on the  GPU partition and allocated ten nodes, with four NVIDIA A100s per node.  NMFk with Binary Bleed Early Stop and standard ran on  $\mathcal{K}=\{2,3,\cdots,100\}$. Early stop visited 60\% of the total compared to Standard NMFk. Both agreed the $k_{optimal}=71$ for the vocabulary size 10,280.

\subsection{Distributed Setting}

To showcase the efficacy of our approach on the largest dataset, we utilized the distributed NMF framework pyDNMFk results and the distributed RESCAL framework pyDRESCAL, as referenced in \cite{pyDNMFk} and \cite{BHATTARAI2023104709}, respectively. Large datasets need substantial computational resources to factorize. For instance, without Binary Bleed, pyDNMFk required 2 hours with 52,000 cores to estimate $k$ using standard NMF for a 50TB dataset, averaging 17.14 minutes per $k$ for $\mathcal{K}=\{2,3,\cdots,8\}$. Similarly, pyDRESCALk required 3 hours with 4,096 cores to factorize 11.5TB of synthetic data using standard RESCAL, averaging 18 minutes per $k$ for $\mathcal{K}=\{2,3,\cdots,11\}$. Binary Bleed Vanilla and Early Stop results on this distributed data were identical, so only the former's results were considered, given that the stop thresholds were crossed on the last $k$. For both RESCAL and NMFk, the selected $k$ matched the standard.

\textbf{RESCAL}: Binary Bleed was applied to the silhouette and relative error metrics with Pre-order and Post-order traversal. 
\begin{wrapfigure}{htb}{0.5\columnwidth}
     \includegraphics[width=.5\columnwidth]{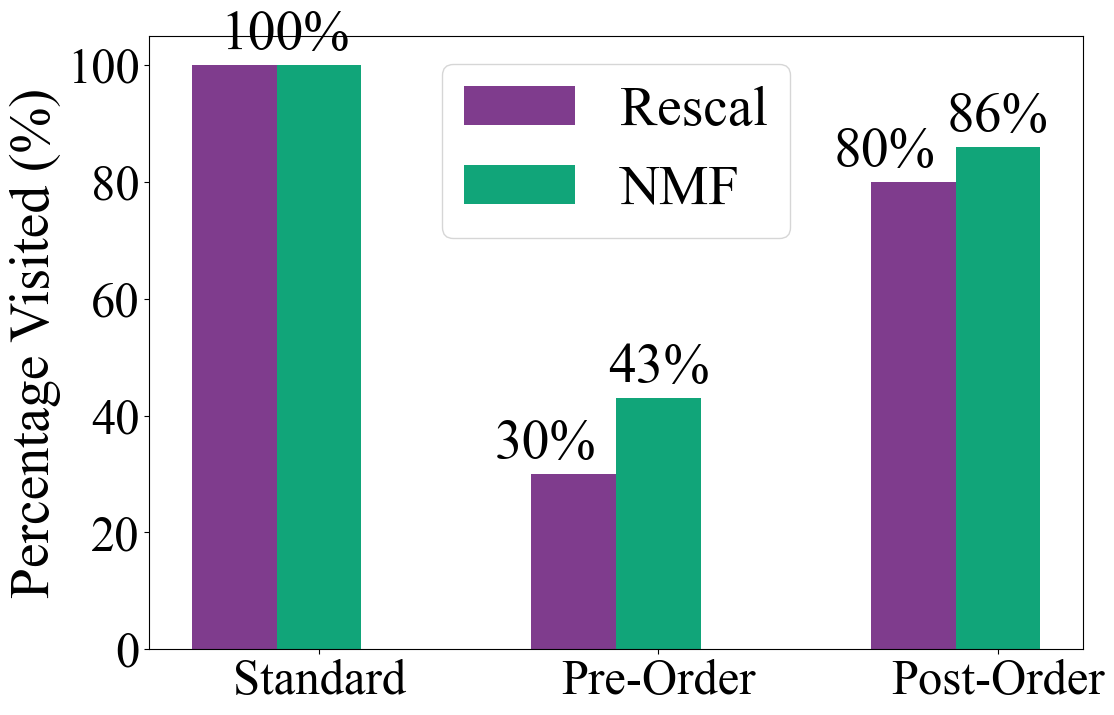}
    \caption{ Binary Bleed reduction on distributed Rescal (Purple), and distributed NMF (Green).}
        \label{fig:distributed_results}
\end{wrapfigure}
 For Pre-order traversal, 30\% of the total $k$ values were visited, resulting in an average runtime of 54 minutes, compared to the 180 minutes required for the standard RESCAL. In contrast, Post-Order traversal visited 80\% of the total $k$ values, with an average runtime of 144 minutes, as in Figure~\ref{fig:distributed_results}. 

\textbf{NMF:} Similarly, the results from \cite{pyDNMFk} were analyzed for $\mathcal{K}=\{2,3,\cdots,8\}$. Using Pre-order and post-order traversal, we evaluated the Binary Bleed Vanilla algorithm for the silhouette score. For Pre-Order traversal, 43\% of the total $k$ values were visited, resulting in an average runtime of 51.43 minutes, compared to the 120 minutes required for the standard NMF. Post-order traversal visited 86\% of the total $k$ values, with an average runtime of 102.86 minutes, as shown in Figure~\ref{fig:distributed_results}. These results further emphasize the efficiency of the proposed approach in reducing computational time while ensuring effective factorization.

\section{Conclusion}
\label{sec:conclusion}
This work addresses the computationally expensive task of $k$ search over potentially large datasets, reducing a $O(n)$ search time to approach $O(log(n))$. Introduced is the ability to optimize the $k$ search from calculation outputs in an upward or lower direction based on a threshold. The algorithm can operate on a single thread, multiple threads, multiple nodes, and distributed systems.  Additionally, early stopping is introduced to reduce directional $k$ bleed on sufficient domains. We tested our method with NMFk, K-means, and RESCAL with Silhouette and Davies Boulding scoring techniques both on synthetic and real-world data. Our experiments showed the $k$ search space can be drastically reduced with Binary Bleed.

\section*{Acknowledgment}
This research used resources provided by the LANL Institutional Computing Program supported by the U.S. Department of Energy National Nuclear Security Administration under Contract No. 89233218CNA000001.

\bibliographystyle{IEEEtran}
\bibliography{references.bib}

\begin{thebibliography}{10}
\providecommand{\url}[1]{#1}
\csname url@samestyle\endcsname
\providecommand{\newblock}{\relax}
\providecommand{\bibinfo}[2]{#2}
\providecommand{\BIBentrySTDinterwordspacing}{\spaceskip=0pt\relax}
\providecommand{\BIBentryALTinterwordstretchfactor}{4}
\providecommand{\BIBentryALTinterwordspacing}{\spaceskip=\fontdimen2\font plus
\BIBentryALTinterwordstretchfactor\fontdimen3\font minus \fontdimen4\font\relax}
\providecommand{\BIBforeignlanguage}[2]{{%
\expandafter\ifx\csname l@#1\endcsname\relax
\typeout{** WARNING: IEEEtran.bst: No hyphenation pattern has been}%
\typeout{** loaded for the language `#1'. Using the pattern for}%
\typeout{** the default language instead.}%
\else
\language=\csname l@#1\endcsname
\fi
#2}}
\providecommand{\BIBdecl}{\relax}
\BIBdecl

\bibitem{alexandrov2020patent}
B.~Alexandrov, L.~Alexandrov, and V.~S. et~al., ``Source identification by non-negative matrix factorization combined with semi-supervised clustering,'' \emph{US Patent S10,776,718}, 2020.

\bibitem{SmartTensors}
\BIBentryALTinterwordspacing
B.~Alexandrov, V.~Vesselinov, and K.~O. Rasmussen, ``Smarttensors unsupervised ai platform for big-data analytics,'' Los Alamos National Lab.(LANL), Los Alamos, NM (United States), Tech. Rep., 2021, lA-UR-21-25064. [Online]. Available: \url{{https://www.lanl.gov/collaboration/smart-tensors/}}
\BIBentrySTDinterwordspacing

\bibitem{TELF}
\BIBentryALTinterwordspacing
M.~Eren, N.~Solovyev, R.~Barron, M.~Bhattarai, D.~Truong, I.~Boureima, E.~Skau, K.~Rasmussen, and B.~Alexandrov, ``{Tensor Extraction of Latent Features (T-ELF)},'' {Los Alamos National Laboratories}, Tech. Rep., Oct. 2023. [Online]. Available: \url{https://github.com/lanl/T-ELF}
\BIBentrySTDinterwordspacing

\bibitem{bhattarai2023distributed}
M.~Bhattarai, I.~Boureima, E.~Skau, B.~Nebgen, H.~Djidjev, S.~Rajopadhye, J.~P. Smith, B.~Alexandrov \emph{et~al.}, ``Distributed non-negative rescal with automatic model selection for exascale data,'' \emph{Journal of Parallel and Distributed Computing}, vol. 179, p. 104709, 2023.

\bibitem{boureima2022distributed}
I.~Boureima, M.~Bhattarai, M.~E. Eren, N.~Solovyev, H.~Djidjev, and B.~S. Alexandrov, ``Distributed out-of-memory svd on cpu/gpu architectures,'' in \emph{2022 IEEE High Performance Extreme Computing Conference (HPEC)}.\hskip 1em plus 0.5em minus 0.4em\relax IEEE, 2022, pp. 1--8.

\bibitem{boureima2024distributed}
I.~Boureima, M.~Bhattarai, M.~Eren, E.~Skau, P.~Romero, S.~Eidenbenz, and B.~Alexandrov, ``Distributed out-of-memory nmf on cpu/gpu architectures,'' \emph{The Journal of Supercomputing}, vol.~80, no.~3, pp. 3970--3999, 2024.

\bibitem{bhattarai2020distributed}
M.~Bhattarai, G.~Chennupati, E.~Skau, R.~Vangara, H.~Djidjev, and B.~S. Alexandrov, ``Distributed non-negative tensor train decomposition,'' in \emph{2020 IEEE High Performance Extreme Computing Conference (HPEC)}.\hskip 1em plus 0.5em minus 0.4em\relax IEEE, 2020, pp. 1--10.

\bibitem{BHATTARAI2023104709}
\BIBentryALTinterwordspacing
M.~Bhattarai, N.~kharat, I.~Boureima, E.~Skau, B.~Nebgen, H.~Djidjev, S.~Rajopadhye, J.~P. Smith, and B.~Alexandrov, ``Distributed non-negative rescal with automatic model selection for exascale data,'' \emph{Journal of Parallel and Distributed Computing}, vol. 179, p. 104709, 2023. [Online]. Available: \url{https://www.sciencedirect.com/science/article/pii/S0743731523000710}
\BIBentrySTDinterwordspacing

\bibitem{10.1145/3558100.3563844}
\BIBentryALTinterwordspacing
M.~E. Eren, N.~Solovyev, M.~Bhattarai, K.~O. Rasmussen, C.~Nicholas, and B.~S. Alexandrov, ``Senmfk-split: Large corpora topic modeling by semantic non-negative matrix factorization with automatic model selection,'' in \emph{Proceedings of the 22nd ACM Symposium on Document Engineering}, ser. DocEng '22.\hskip 1em plus 0.5em minus 0.4em\relax New York, NY, USA: Association for Computing Machinery, 2022. [Online]. Available: \url{https://doi.org/10.1145/3558100.3563844}
\BIBentrySTDinterwordspacing

\bibitem{10527250}
M.~E. Eren, R.~Barron, M.~Bhattarai, S.~Wanna, N.~Solovyev, K.~Rasmussen, B.~S. Alcxandrov, and C.~Nicholas, ``Catch'em all: Classification of rare, prominent, and novel malware families,'' in \emph{2024 12th International Symposium on Digital Forensics and Security (ISDFS)}, 2024, pp. 1--6.

\bibitem{10.1145/3624567}
\BIBentryALTinterwordspacing
M.~E. Eren, M.~Bhattarai, R.~J. Joyce, E.~Raff, C.~Nicholas, and B.~S. Alexandrov, ``Semi-supervised classification of malware families under extreme class imbalance via hierarchical non-negative matrix factorization with automatic model selection,'' \emph{ACM Trans. Priv. Secur.}, sep 2023, just Accepted. [Online]. Available: \url{https://doi.org/10.1145/3624567}
\BIBentrySTDinterwordspacing

\bibitem{MacQueen1967SomeMF}
\BIBentryALTinterwordspacing
J.~MacQueen, ``Some methods for classification and analysis of multivariate observations,'' 1967. [Online]. Available: \url{https://api.semanticscholar.org/CorpusID:6278891}
\BIBentrySTDinterwordspacing

\bibitem{medioids}
\BIBentryALTinterwordspacing
\emph{Partitioning Around Medoids (Program PAM)}.\hskip 1em plus 0.5em minus 0.4em\relax John Wiley and Sons, Ltd, 1990, ch.~2, pp. 68--125. [Online]. Available: \url{https://onlinelibrary.wiley.com/doi/abs/10.1002/9780470316801.ch2}
\BIBentrySTDinterwordspacing

\bibitem{jain1988algorithms}
A.~K. Jain and R.~C. Dubes, \emph{Algorithms for clustering data}.\hskip 1em plus 0.5em minus 0.4em\relax Prentice-Hall, Inc., 1988.

\bibitem{BEZDEK1984191}
\BIBentryALTinterwordspacing
J.~C. Bezdek, R.~Ehrlich, and W.~Full, ``Fcm: The fuzzy c-means clustering algorithm,'' \emph{Computers \& Geosciences}, vol.~10, no.~2, pp. 191--203, 1984. [Online]. Available: \url{https://www.sciencedirect.com/science/article/pii/0098300484900207}
\BIBentrySTDinterwordspacing

\bibitem{mini_batch}
\BIBentryALTinterwordspacing
D.~Sculley, ``Web-scale k-means clustering,'' in \emph{Proceedings of the 19th International Conference on World Wide Web}, ser. WWW '10.\hskip 1em plus 0.5em minus 0.4em\relax New York, NY, USA: Association for Computing Machinery, 2010, p. 1177–1178. [Online]. Available: \url{https://doi.org/10.1145/1772690.1772862}
\BIBentrySTDinterwordspacing

\bibitem{spherical_k_means}
K.~Hornik, I.~Feinerer, M.~Kober, and C.~Buchta, ``Spherical k-means clustering,'' \emph{Journal of Statistical Software}, vol.~50, pp. 1--22, 09 2012.

\bibitem{elkan2003using}
C.~Elkan, ``Using the triangle inequality to accelerate k-means,'' in \emph{Proceedings of the 20th international conference on Machine Learning (ICML-03)}, 2003, pp. 147--153.

\bibitem{xu2003document}
W.~Xu, X.~Liu, and Y.~Gong, ``Document clustering based on non-negative matrix factorization,'' in \emph{Proceedings of the 26th annual international ACM SIGIR conference on Research and development in informaion retrieval}, 2003, pp. 267--273.

\bibitem{symmetric_nmf}
R.~Vangara, K.~Rasmussen, G.~Chennupati, and B.~Alexandrov, ``Determination of the number of clusters by symmetric non-negative matrix factorization,'' in \emph{SPIE}, 04 2021, p.~15.

\bibitem{nickel2011three}
M.~Nickel, V.~Tresp, H.-P. Kriegel \emph{et~al.}, ``A three-way model for collective learning on multi-relational data.'' in \emph{Icml}, vol.~11, no. 10.5555, 2011, pp. 3\,104\,482--3\,104\,584.

\bibitem{Karp2007NoisyBS}
\BIBentryALTinterwordspacing
R.~M. Karp and R.~D. Kleinberg, ``Noisy binary search and its applications,'' in \emph{ACM-SIAM Symposium on Discrete Algorithms}, 2007. [Online]. Available: \url{https://api.semanticscholar.org/CorpusID:774118}
\BIBentrySTDinterwordspacing

\bibitem{Tyrrell1988}
\BIBentryALTinterwordspacing
R.~A. Tyrrell and D.~A. Owens, ``A rapid technique to assess the resting states of the eyes and other threshold phenomena: The modified binary search (mobs),'' \emph{Behavior Research Methods, Instruments, \& Computers}, vol.~20, no.~2, pp. 137--141, 1988. [Online]. Available: \url{https://doi.org/10.3758/BF03203817}
\BIBentrySTDinterwordspacing

\bibitem{ChadhaMM14}
\BIBentryALTinterwordspacing
A.~R. Chadha, R.~Misal, and T.~Mokashi, ``Modified binary search algorithm,'' \emph{CoRR}, vol. abs/1406.1677, 2014. [Online]. Available: \url{http://arxiv.org/abs/1406.1677}
\BIBentrySTDinterwordspacing

\bibitem{efficient_parallel_binary}
D.~Chen, ``Efficient parallel binary search on sorted arrays, with applications,'' \emph{IEEE Transactions on Parallel and Distributed Systems}, vol.~6, no.~4, pp. 440--445, 1995.

\bibitem{parallel_binary}
S.~Akl and H.~Meijer, ``Parallel binary search,'' \emph{IEEE Transactions on Parallel Distributed Systems}, vol.~1, no.~02, pp. 247--250, apr 1990.

\bibitem{pyDNMFk}
M.~Bhattarai, B.~Nebgen, E.~Skau, M.~Eren, G.~Chennupati, R.~Vangara, H.~Djidjev, J.~Patchett, J.~Ahrens, and B.~ALexandrov, ``pydnmfk: Python distributed non negative matrix factorization,'' \url{https://github.com/lanl/pyDNMFk}, 2021.

\bibitem{distributed_constrained_heuristic_search}
K.~Sycara, S.~Roth, N.~Sadeh, and M.~Fox, ``Distributed constrained heuristic search,'' \emph{IEEE Transactions on Systems, Man, and Cybernetics}, vol.~21, no.~6, pp. 1446--1461, 1991.

\bibitem{distributedNearestNeighbor}
X.~Liu, Z.~Li, C.~Deng, and D.~Tao, ``Distributed adaptive binary quantization for fast nearest neighbor search,'' \emph{IEEE Transactions on Image Processing}, vol.~26, no.~11, pp. 5324--5336, 2017.

\bibitem{big_data_knn}
\BIBentryALTinterwordspacing
A.~B. Hassanat, ``Furthest-pair-based binary search tree for speeding big data classification using k-nearest neighbors,'' \emph{Big Data}, vol.~6, no.~3, pp. 225--235, 2018. [Online]. Available: \url{https://doi.org/10.1089/big.2018.0064}
\BIBentrySTDinterwordspacing

\bibitem{sismanis2012parallel}
N.~Sismanis, N.~Pitsianis, and X.~Sun, ``Parallel search of k-nearest neighbors with synchronous operations,'' in \emph{2012 IEEE Conference on High Performance Extreme Computing}.\hskip 1em plus 0.5em minus 0.4em\relax IEEE, 2012, pp. 1--6.

\bibitem{cormen2022introduction}
T.~H. Cormen, C.~E. Leiserson, R.~L. Rivest, and C.~Stein, \emph{Introduction to Algorithms}, 4th~ed.\hskip 1em plus 0.5em minus 0.4em\relax MIT Press, 2022.

\bibitem{barron2024cybersecurity}
R.~Barron, M.~E. Eren, M.~Bhattarai, S.~Wanna, N.~Solovyev, K.~Rasmussen, B.~S. Alexandrov, C.~Nicholas, and C.~Matuszek, ``Cyber-security knowledge graph generation by hierarchical nonnegative matrix factorization,'' 2024.

\end{thebibliography}

\end{document}